\documentclass{rsproca}

\usepackage{graphicx}
\usepackage{amsmath}
\usepackage{amsfonts}
\usepackage{bm}
\usepackage{bbm}
\usepackage{url}
\usepackage{amssymb}
\usepackage{mathrsfs}
\usepackage{dsfont}
\usepackage{subfigure}
\usepackage{paralist}
\usepackage{xcolor}

\usepackage{stmaryrd}

\newcommand{\zero}{\bm{0}}
\newcommand{\vel}{\bm{v}}
\renewcommand{\acc}{\bm{a}}
\newcommand{\n}{\bm{n}}
\newcommand{\ut}{\bm{t}}
\newcommand{\uv}{\bm{u}}
\newcommand{\tp}{\bm{t}^+}
\newcommand{\tm}{\bm{t}^-}

\newcommand{\ex}{\bm{e}_x}
\newcommand{\ey}{\bm{e}_y}
\newcommand{\jump}[1]{\mbox{$\llbracket #1\rrbracket$}}
\newcommand{\tzero}{\vartheta_0}
\newcommand{\tone}{\vartheta_1}
\newcommand{\vt}{\vartheta}
\newcommand{\lv}{\lambda\vel}
\newcommand{\nua}{\nu^\ast}
\newcommand{\lvtwo}{\lambda v^2}
\newcommand{\lvthree}{\lambda v^3}
\newcommand{\Ws}{W_\mathrm{s}}

\newcommand{\Wg}{W_\mathrm{g}}
\newcommand{\taut}{\tau\ut}
\newcommand{\tautnot}{\tau_0\ut_0}
\newcommand{\sspeed}{\dot{s}_0}
\newcommand{\force}{\bm{f}}
\newcommand{\F}{\bm{F}}

\newcommand{\Neptune}{\Psi}
\newcommand{\Force}{\bm{\Phi}}

\newcommand{\MX}{\emph{Myxococcus xanthus}\ }
\newcommand{\MdX}{\emph{M. xanthus}\ }
\newcommand{\Ncs}{\emph{Nostocales}\ }
\newcommand{\Nc}{\emph{Nostoc}\ }

\newcommand{\nigh}[1]{{#1}}
\newcommand{\nnot}[1]{\left({#1}\right)_0}

\begin{document}

\title{Dissipative Shocks behind Bacteria Gliding}

\author{
Epifanio G. Virga}

\address{Dipartimento di Matematica, Universit\`a di Pavia, Via Ferrata 5, I-27100 Pavia, Italy}

\subject{87.17.Jj, 87.17.Rt, 87.10.Ca, 46.70.Hg}

\keywords{Gliding bacteria, Dissipative shocks, One-dimensional continua, }

\corres{EG Virga\\
\email{eg.virga@unipv.it}}

\begin{abstract}
\emph{Gliding} is a means of locomotion on rigid substrates utilized by a number of bacteria including myxobacteria and cyanobacteria. One of the hypotheses advanced to explain this motility mechanism hinges on the role played by the slime filaments continuously extruded from gliding bacteria. This paper solves in full a non-linear mechanical theory that treats as \emph{dissipative shocks} both the point where the extruded slime filament comes in contact with the substrate, called the filament's \emph{foot}, and the pore on the bacterium outer surface from where the filament is ejected. We prove that kinematic compatibility for shock propagation requires that the bacterium uniform gliding velocity (relative to the substrate) and the slime ejecting velocity (relative to the bacterium) must  be equal, a coincidence that seems to have already been observed.
\end{abstract}


\begin{fmtext}
\section{Introduction}\label{sec:intro}
For living cells, motility is as essential to their life as are the nutrients that sustain it. If food becomes scarce in the vicinity of a cell, this must move to survive. As neatly explained in Wolgemuth's recent review \cite{wolgemuth:biomechanics}, cell motion involves interaction with the environment: to move, a cell must exert an \emph{active}  force on its environment, which is counteracted by a \emph{resistive} force, or reaction. At the cellular length scale inertia is negligible, hence a motion results only from the balance of the active force produced by the cell and the reactive force exerted by the environment. A cell that stops producing a propelling force is brought to a halt in virtually no time.

One way of classifying cellular motion is by distinguishing the environments against which a cell moves into two gross categories: fluids and solids. In the former case, a cell is a \emph{swimmer}, as it moves surrounded by a resistive fluid, in the latter case, it is a \emph{surfer}, as it moves about a rigid surface. Like most gross classifications, also this joins the advantage of simplicity to the disadvantage of inexactness. According to \cite{nan:uncovering}, what we called a \emph{surfer} actually may exhibit one or
more of the following different motility patterns, namely, \emph{swarming},
\end{fmtext}
\maketitle
\noindent\emph{twitching}, \emph{sliding}, and \emph{gliding}.
This paper will only be concerned with the latter.
We shall adopt here the neat definition of gliding given by Jarosch~\cite{jarosch:gliding} in 1962 (also reproduced in \cite{walsby:mucilage}):
\begin{quote}
\emph{Gliding} is ``the active movement of an organism in contact with a solid substratum where there is neither a visible organ responsible for the movement nor a distinct change in the shape of the organism.''\footnote{A perhaps more concise, but equivalent definition is given in \cite{nan:uncovering}: Gliding motility is ``the active and smooth translocation of cells on a surface without the aid of flagella or pili''. Finally, a similar definition taken from \cite{reichenbach:taxonomy} is also utilized in \cite{wolgemuth:biomechanics}, that is, ``the translocation in the long axis of the bacterium when in contact with a surface'' .}
\end{quote}

As remarked by Fritsch~\cite{fritsch:structure} in 1945, gliding movements are usually associated with the secretion of mucilage (also called \emph{slime}), which has often been regarded as the very cause of movement \cite{walsby:mucilage}. Here we are mainly concerned with modeling quantitatively the possible connection between slime extrusion and thrust force on the bacterium that secretes it. It should be clear from the start that the mechanism that we shall envisage is different from that of jet propulsion in the absence of inertia that has recently been studied in \cite{spagnolie:jet} for swimming bacteria: in this view, it is not the slime extrusion that generates the propelling thrust, but the adhesion of the secreted slime on the rigid substrate that sustains the advancing motion. We shall return later to the gliding mechanism modeled in this paper, after having briefly reviewed the families of bacteria to which it could be potentially applied.

Many bacteria glide over surfaces, including cyanobacteria and myxobacteria \cite{burchard:gliding_prokaryotes}. The former are among the fastest gliders, with a velocity as high as $10\ \mu\mathrm{m}\,\mathrm{s}^{-1}$ \cite{reichenbach:taxonomy}, whereas the latter are among the slowest, with \MX going at the most at a speed of $4\ \mu\mathrm{m}\,\mathrm{min}^{-1}$, $1000$ slower than typical flagellated bacteria \cite{hoiczyk:gliding,spormann:gliding,mauriello:gliding}.

\MdX is presumably the most studied among gliding bacteria for the variety of propulsion mechanisms it exhibits and the complexity of its life cycle. It is a common Gram-negative bacterium that lives in the soil \cite{koch:social,shimkets:social}; it is rod-shaped with typical length and diameter of $5$-$7\ \mu\mathrm{m}$ and $0.5\ \mu\mathrm{m}$, respectively. \MdX makes use of two genetically distinct mechanisms for gliding; one, the other, or both may be present in an individual cell. One mechanism, called \emph{social}, is characteristic of synergetic, coordinated motions involving a colony of cells \cite{kaiser:coupling,igoshin:waves,bradley:function}, while the other, called \emph{adventurous}, is typical of single cells living at the outskirts of a colony \cite{hoiczyk:genetics}. What distinguishes social from adventurous gliding in \MdX is not just the collective character of the former, as opposed to the solipsism of the latter. They also differ in their mechanics. Extrusion and retraction of pili are involved in social gliding: each cell emits a filament, typically of nearly $6\ \mathrm{nm}$ in diameter and a few microns in length, which adheres to either the supporting substrate or to another cell nearby; by retracting the adhered filament, a cell propels itself in coordination with those in its vicinity.

\begin{figure}[!h]
  \centering
  \subfigure[]{\includegraphics[width=.4\linewidth]{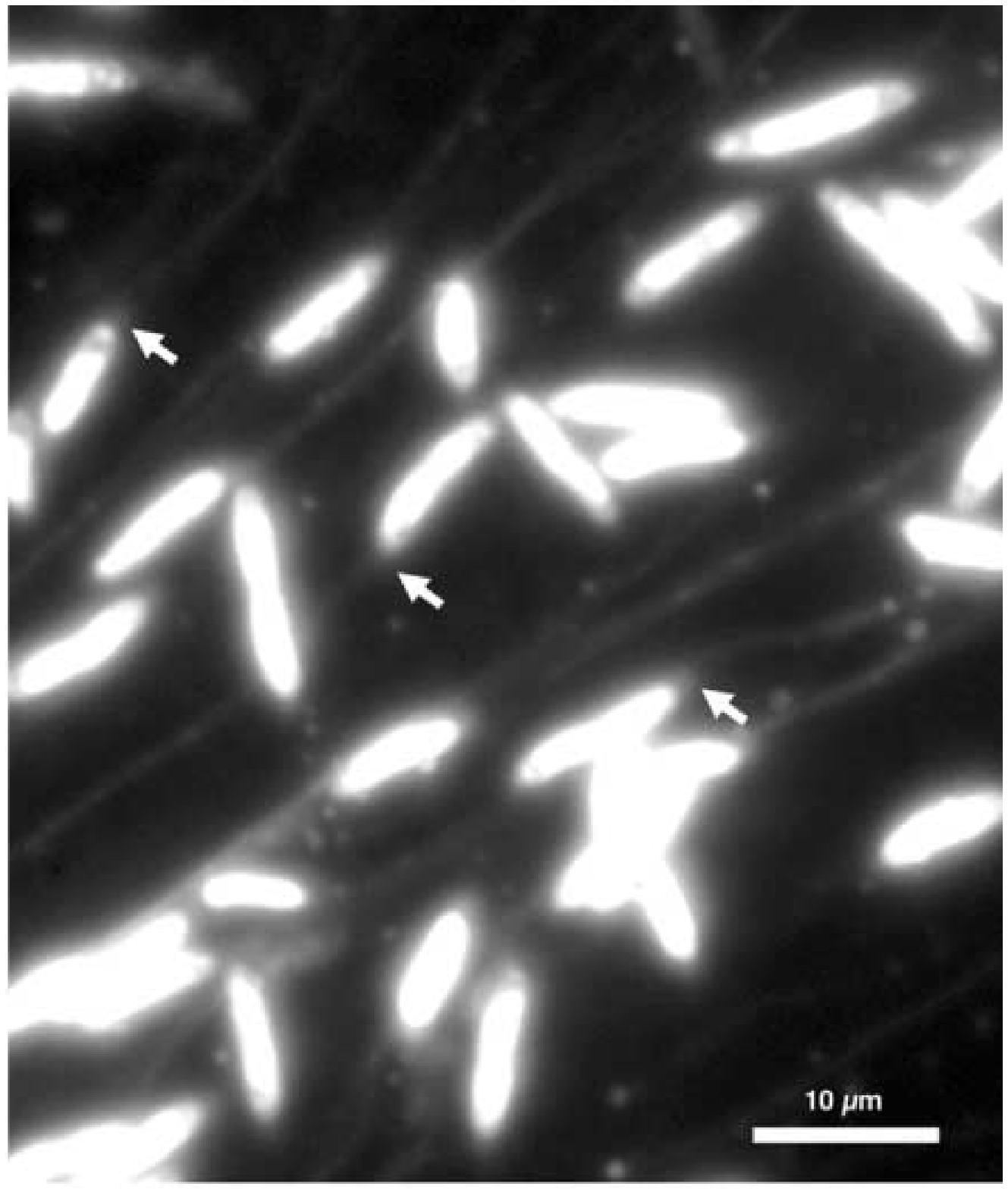}}
  \hspace{.05\linewidth}
  \subfigure[]{\includegraphics[width=.39\linewidth]{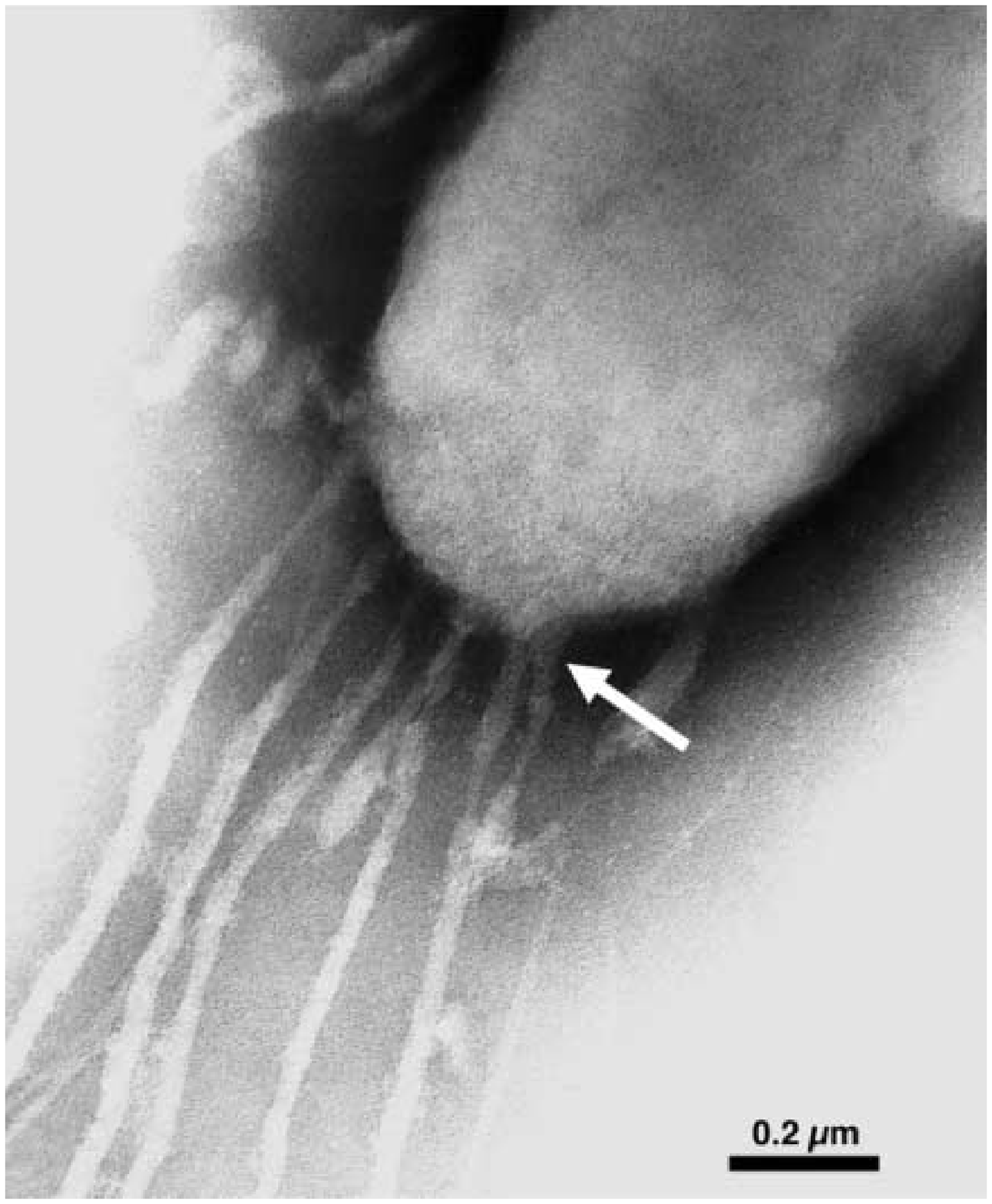}}
  \caption{
  Examination of the slime secretion process in Wild-Type \MX cells.
  (a) Fluorescent light micrograph of gliding \MdX cells (capable of both adventurous and social gliding: Strain DK1622). During locomotion, the cells leave slime trails behind, which can be stained by Acridine orange. The slime trails originate at the rear poles of the individual cells (small arrows). Photograph taken after $1\ \mathrm{hr}$ at $2000\times$.
  (b) Electron micrograph of the cell pole of a gliding \MdX cell. At higher magnification, it can be seen that the slime trails are composed of several slime bands, which are secreted from the sites at the cell pole, where the nozzles are located (large arrow).
  This figure is reproduced with permission from \cite{wolgemuth:how}, Copyright Elsevier (2002) (Licence number $3347720296046$, granted on the 14th of March 2014).}
\label{fig:MX_micrograph}
\end{figure}

If social gliding of \MdX is rather well understood, its adventurous gliding remains a mystery. Apart from not being powered by pili, little is known about the actual mechanism of adventurous gliding of myxobacteria, as none of the hypotheses advanced so far has proven able to explain the full body of available experimental evidence. One such hypothesis, put forward nearly 90 years ago \cite{jahn:polyangiden,kuhlwein:weitere}, hinges on the observation that myxobacteria secrete slime \cite{stanier:elasticotaxis,fontes:myxococcus}. More recently, Wolgemuth \emph{et al.}~\cite{wolgemuth:how} gave direct evidence that slime extrusion is associated with the adventurous gliding of \MdX: the micrographs in Fig.~\ref{fig:MX_micrograph},
which lately have been repeatedly reproduced, are rather striking illustrations of the slime filaments ejected by the cell. A mathematical model was also proposed in \cite{wolgemuth:how} to describe the extrusion of slime from the cell's nozzles; it treated it as a polyelectrolyte gel and assumed that its extrusion resulted from the swelling induced by a hydration process taking place within the nozzle's body. This rather sophisticated jet emission mechanism allowed for an estimate of the propelling force, which was found sufficient to explain the observed gliding velocity both for myxobacteria and cyanobacteria.\footnote{See also \cite{wolgemuth:junctional} for the explanation of their experimental findings within this theoretical framework.}

An opposing hypothesis has recently been advanced, according to which adventurous motility is provided by protein motors distributed along the length of the cell, which tend to form complexes that remain stationary relative to the supporting rigid substrate while the bacterium moves \cite{sun:effect,sliusarenko:motors,mignot:evidence,mauriello:gliding,luciano:emergence}. Such \emph{focal adhesion} complexes seem to require slime secreted underneath the cell both to produce specific adhesion sites and to lubricate the area of contact between cell and substrate, as also confirmed by some other recent observations \cite{ducret:wet}. Also compromising hypotheses have been proposed to the effect of regarding both slime extrusion and focal adhesion as viable coexisting mechanisms for myxobacteria gliding. No wonder if in such a state of affairs controversies have grown \cite{nan:uncovering}. One, in particular, concerned the necessity for a gliding cell to rotate along its long axis while advancing: this was first believed to be the case for the focal adhesion theory to be confirmed, as the \MdX cytoskeleton along which the motive proteins are supposed to move is helicoidal with axis coincident with the cell's long axis \cite{mauriello:gliding,jarrell:surprisingly}. Very recently, evidence has been brought against cell rotation \cite{koch:characterization}, though \emph{per se} this would neither rule out completely the focal adhesion hypothesis \cite{nan:flagella}, nor prove the slime extrusion hypothesis.

As controversial as this latter hypothesis may be to explain in full the gliding of myxobacteria, since the work of Hoiczyk and Baumeister \cite{hodgkin:junctional}, it is well established as an explanation for the gliding of cyanobacteria. These latter are filamentous bacteria (blue-green algae) with a number of septa, for which the slime extrusion hypothesis assumes that slime filaments are ejected through the pores that surround each cell septum (see also \cite{hoiczyk:gliding}). Hoiczyk and Baumeister showed in \cite{hodgkin:junctional} how slime formed bands about the cell surface that could be removed by a fluid flow. They also were able to establish that the slime was emanated at a rate of $3\ \mu\mathrm{m}\,\mathrm{s}^{-1}$, which compared well with the gliding velocity (a coincidence to which we shall return later).

\begin{figure}[!h]
  \centering
  \subfigure[]{\includegraphics[width=.4\linewidth]{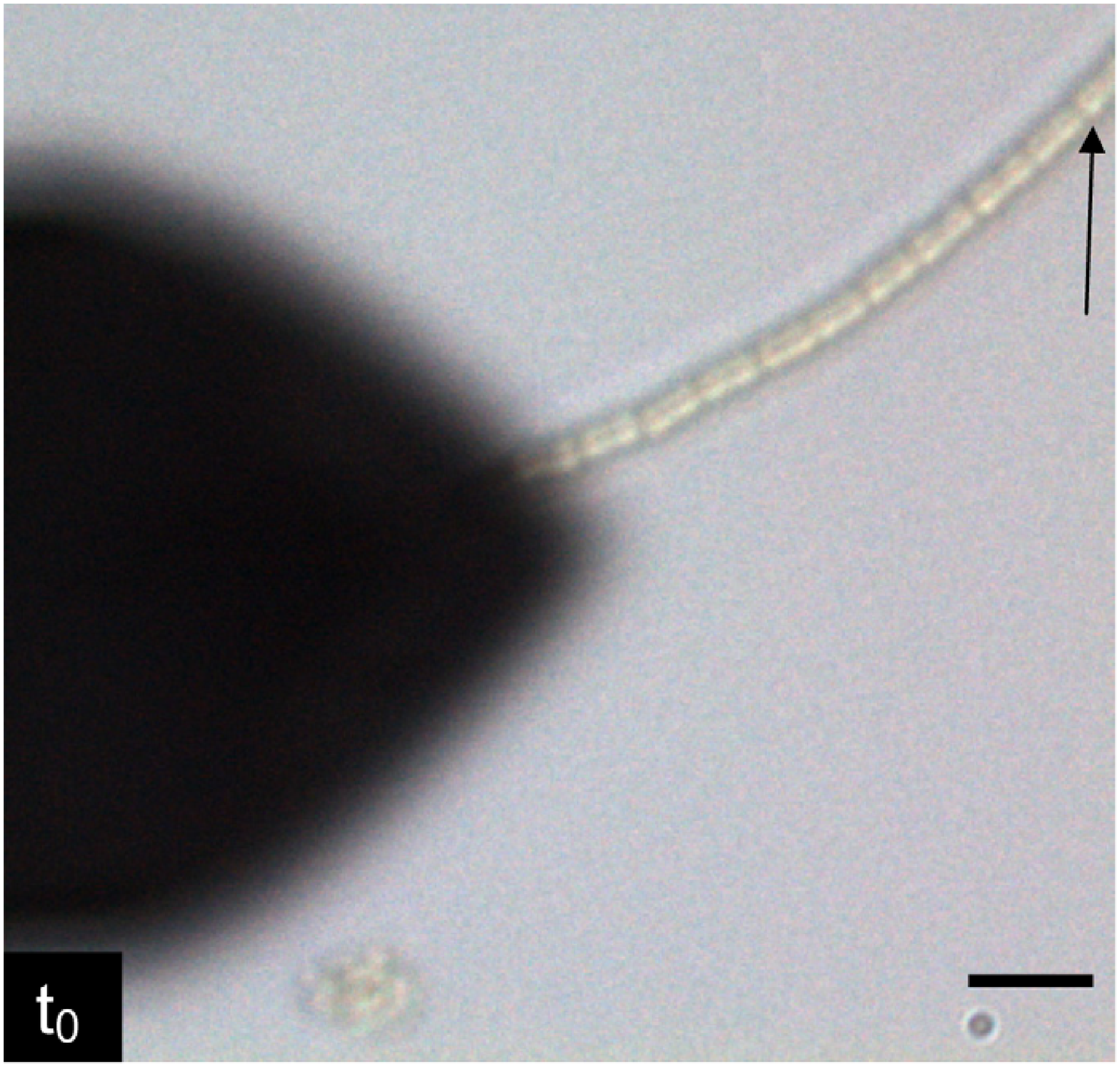}}
  \hspace{.05\linewidth}
  \subfigure[]{\includegraphics[width=.385\linewidth]{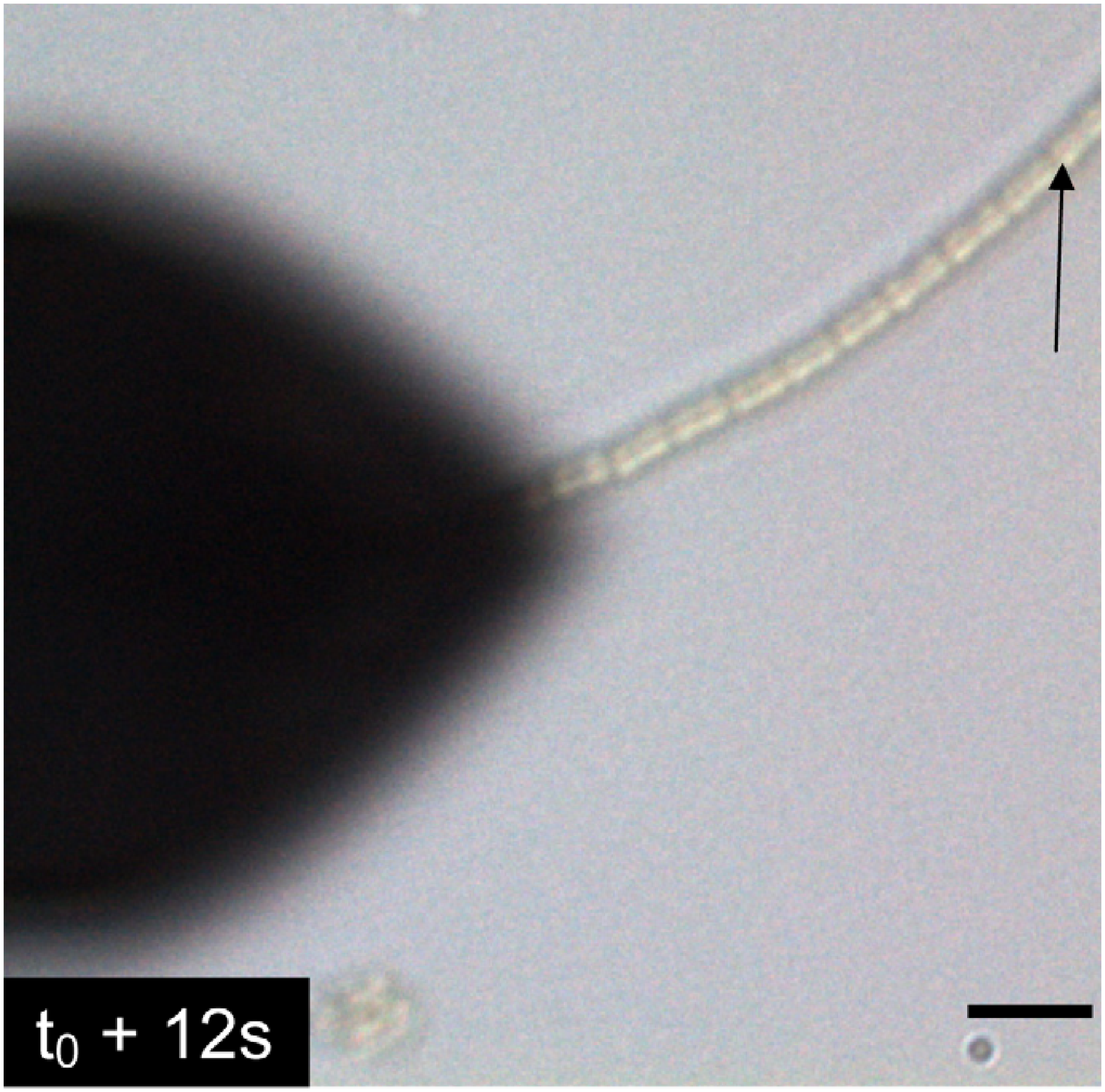}}
  \caption{
Optical snapshots of a  \Nc bacterium gliding upon a glass slide. The scale is given by the black reference bar (corresponding to $10\ \mu\mathrm{m}$). The bacterium is moving from right to left as indicated by the displacement of the arrow landmark. Image (b) is taken $12\ \mathrm{s}$ after image (a). Both images are reproduced from \cite{dhahri:in-situ}. It should be noted that the resolution of these images is at least 100 times poorer than that of the images shown in Fig.~\ref{fig:MX_micrograph}, and so the slime filaments cannot be seen.}
\label{fig:nostoc}
\end{figure}

A previous study had shown the distribution of pores ($14$-$16\ \mathrm{nm}$ in diameter) close to the cyanobacteria septa \cite{guglielmi:structure}. These pores were recognized in \cite{hodgkin:junctional} as parts of a more complex system (called \emph{junctional pore complex}) which appeared to constitute the extrusion organelles \cite{hoiczyk:envelope}.
Figure~\ref{fig:nostoc} presents recent optical images illustrating the gliding of a \Nc cyanobacterium, belonging to the \Ncs\!, a group that contains most of the cyanobacteria capable of gliding \cite{dhahri:in-situ}.\footnote{This work was actually only marginally concerned with optical observations; its main focus was on the in-situ study of the slime secretion by means of AFM.}

The general mathematical model put forward in \cite{wolgemuth:how} builds on the assumption that external fluids perfuse into the nozzle-like organelles hydrating the polysaccharide material constituting the slime; the swelling thus produced forces the slime out of the pores and causes the swollen gel to fall on the substrate and adhere to it. This, as it were, produces a footing for the cell to advance in a way that only superficially resembles walking \cite{wolgemuth:biomechanics}. Though the extrusion mechanism and the force associated with the gel swelling are fairly well described by the model in \cite{wolgemuth:how}, to my knowledge, the forces associated with both adhesion and footing have not yet been studied. In particular, it would be desirable to know how the propelling force would be determined by the full gliding mechanism, which includes slime extrusion, adhesion, and footing.

\begin{figure}[!h]
\centering
  \includegraphics[width=.4\linewidth]{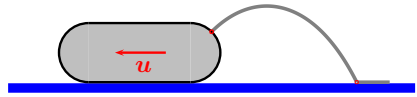}
  \caption{
  \nigh{Cartoon of a cigar-shaped bacterium gliding with constant velocity $\uv$; both its pore and its foot are represented as thick (red) points.}
  }
\label{fig:cartoon}
\end{figure}

This will be the subject of this paper, which builds on a theory that has recently been proven useful to explain the forces at play in the popular, fascinating phenomenon of the \emph{chain fountain} \cite{virga:dissipative}, a common chain that under the action of gravity alone raises spontaneously out of the pot that contains it before plunging down towards the floor \cite{mould:self}.\footnote{A similar phenomenon had indeed been documented earlier by Hanna and King \cite{hanna:instability} (see also \cite{hanna:slack}), but had not met with the popular response of the web that welcomed Mould's movie \cite{mould:self}, which was viewed by millions of people, including the writer of this paper.} The feature of the theory presented in \cite{virga:dissipative} that distinguishes it from the earlier theory of Biggins and Warner \cite{biggins:understanding,biggins:growth} is the assumption that the pickup and putdown points of the steady, shape preserving motion of the chain fountain are standing \emph{shocks} that dissipate energy at the rate dictated by a classical law for internal impacts. Here the same theory is applied to the extruded slime filament, which comes in contact with the substrate over which a bacterium glides. Both the \emph{foot}, where contact is established between filament and substrate, and the \emph{pore}, where the filament is extruded from the bacterium, will be treated as \emph{dissipative shocks}, though of two different natures. \nigh{Figure~\ref{fig:cartoon} illustrates the model: a cigar-shaped bacterium glides on a flat substrate with constant velocity $\uv$ while ejecting a slime filament from its back (actually, the bacterium is apolar and its temporary back is just the end opposite to the direction of its motion).}

In a steady gliding motion, the foot itself glides over the substrate and the filament shape is dislocated in time in the substrate rest frame. However, in the bacterium rest frame, both the filament shape and its foot remain unchanged as time elapses, making the extruded filament completely analogous to the ``ejected'' chain studied in \cite{virga:dissipative}. In this frame, both the ejecting pore and the adhering foot of the filament will be regarded as standing shocks; they will be treated following a paper of O'Reilly and Varadi~\cite{reilly:treatment} who, elaborating on earlier work of Green and Naghdi~\cite{green:thermodynamics,green:note,green:derivation,green:thermal}, proposed an elegant and rather comprehensive theory of shocks in one-dimensional continua, which in Section~\ref{sec:shocks} is extended to the case at hand.

Section~\ref{sec:shocks}, which is largely based on \cite{virga:dissipative} and serves the purpose of making our development here self-contained, is split in several subsections to make it easier for the reader retrace the different elements of the theory developed here. In Section~\ref{sec:gliding}, we shall describe in quantitative terms the solution that the theory recalled in Section~\ref{sec:shocks} affords for a gliding bacterium. In particular, we shall compute both the bacterium gliding velocity and the force resulting on it from the complete motion of the extruded slime filament. One simple result, which has already echoed in the observations recalled above, will follow from a kinematic compatibility condition on shock propagation, that is, that the absolute gliding velocity must equal the relative extrusion velocity.\footnote{The former is taken by an observer at rest on the substrate, whereas the latter is taken by an observer at rest on the gliding bacterium.} \nigh{The analysis in Section~\ref{sec:shocks} does not neglect gravity and in a way treats the ejected slime filament macroscopically, as it were composed of coherent chain links fired away from an orifice. This assumption is discussed in Section~\ref{sec:zero_gravity} together with the consequences of relaxing it.} Finally, the paper is closed by Section~\ref{sec:conclusion}, where we collect the theoretical conclusions reached here and comment about the want for their experimental scrutiny.

\section{Dissipative Shocks}\label{sec:shocks}
In this section we recall the theory presented in \cite{virga:dissipative} adapting it to the special problem envisaged here.
Think of a slime filament as an \emph{inextensible} string with uniform mass density $\lambda$ per unit length, parameterized in the reference configuration by the arc-length $s$. The position in space occupied by a material point of the filament is represented  by the mapping $p=p(s,t)$. Here $s$, which designates the convected variable, could as well be used to designate the arc-length in the present configuration. Correspondingly, the velocity $\vel$ is defined by $\vel:=\dot{p}$, where a superimposed dot represents differentiation with respect to time $t$. Similarly, $\acc:=\dot{\vel}$ is the acceleration. Let $\force$ denote the \emph{external} force acting per unit length of the filament and $\tau\geqq0$ the internal tension that arises as a reaction to the inextensibility constraint. The balance of linear momentum along any smooth arc of the filament is expressed by
\begin{equation}\label{eq:smooth_momentum_balance}
\lambda\acc=\force+(\taut)^\prime,
\end{equation}
where a prime $^\prime$ denotes differentiation with respect to $s$ (see, for example, \cite[Section\,34]{villaggio:mathematical}). Equation \eqref{eq:smooth_momentum_balance} is written for an observer gliding at uniform speed along with the bacterium. \nigh{Though the bacterium gliding motion on the substrate will be assumed to have uniform speed as a result of neglecting its inertia, in describing the filament's motion, with the objective of determining the bacterium thrust, we shall neither neglect the filament's inertia nor the gravity (which we regard somehow on the same footing); we defer the reader to Section~\ref{sec:zero_gravity} for a quantitative discussion of this assumption.}

\subsection{Shock Equations}\label{sec:shock_equations}
\begin{figure}[!h]
\centering
  \includegraphics[width=.3\linewidth]{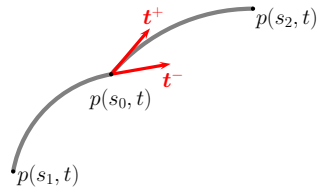}
  \caption{
  The present shape at time $t$ of the filament. The points $p(s_1,t)$ and $p(s_2,t)$, with $s_2>s_1$, delimit the arc under consideration. The point $p(s_0,t)$ is a singular point, where the unit tangent $\ut$ is discontinuous, with traces $\tp$ and $\tm$ on the two sides.}
\label{fig:shock}
\end{figure}

In this context, a \emph{shock} propagating along the filament is described by a function, $s_0=s_0(t)$, identifying the point in the reference configuration carrying a discontinuity in speed. Specifically, we assume that $p$ is continuous at $s_0$, because the ejected filament breaks nowhere, but $\vel$ is discontinuous. Similarly, both the unit tangent $\ut$, the principal unit normal $\n$, and the curvature $c$ of the curve representing the present shape (at time $t$) of the filament are discontinuous at $p(s_0(t),t)$, as illustrated in Fig.~\ref{fig:shock}.
We shall call $p(s_0,t)$ a \emph{singular point}. We denote by $\tp$ and $\tm$ the two limiting values of $\ut$ across a singular point. Here and below, for any quantity $\Neptune$, superscripts $^\pm$ refer to the traces of $\Neptune$ across $p(s_0,t)$ from the sides of increasing and decreasing $s$, respectively. Also, we shall employ the customary notation $\jump{\Neptune}:=\Neptune^+-\Neptune^-$, for the \emph{jump} of $\Neptune$ across a singular point.

The shock speed is $\sspeed$ relative to both the reference and present shapes (as a consequence of the filament's inextensibility). A kinematic compatibility condition arises for the jump of $\vel$, as a result of the requirement that the velocity of the geometric point which instantaneously coincides in space with a singular point can be expressed in two different, but equivalent ways (see, for example, \cite{reilly:treatment,reilly:energetics}):
\begin{equation}\label{eq:kinematic_compatibility_velocity}
\jump{\vel}+\sspeed\jump{\ut}=\zero.
\end{equation}

The balance of linear momentum for an arbitrary small arc enclosing a singular point (that is, for $s_2\to s_0^+$ and $s_1\to s_0^-$ in Fig.~\ref{fig:shock}), requires that
\begin{equation}\label{eq:jump_conditions_linear_momentum}
\jump{\taut}+\sspeed\jump{\lv}+\Force=\zero,
\end{equation}
where $\Force$ is the concentrated supply of momentum that must be provided at a singular point to sustain  the shock. In a similar way (see again \cite{reilly:treatment,reilly:energetics} for more details), the energy balance at a singular point results into the following equation,
\begin{equation}\label{eq:jump_conditions_energy}
\jump{\taut\cdot\vel}+\frac12\lambda\sspeed\jump{v^2}+\Ws=0,
\end{equation}
where $\Ws$ is the concentrated power supply involved in the shock.\footnote{Equation \eqref{eq:jump_conditions_energy} is a specialization to the athermal case treated here of equation (2.7)$_4$ of \cite{reilly:treatment}; what here is denoted $\Ws$ was there denoted $\Phi_E$. For $\Ws<0$, energy is lost in the shock. According to the treatment of \emph{chains} in Sommerfeld's book \cite{sommerfeld:mechanics} (see, pp.\,28--29 and Problem I.7, pp.\,241, 257), the energy loss in chain dynamics is a concept first introduced by Lazare Carnot, the father of Sadi (this latter known for his contributions to the theory of heat), who was a writer on mathematics and mechanics (besides later becoming one of the most loyal of Napoleon's generals). See also \cite[p.\,52]{muller:history}, \cite{steiner:equations} and \cite{wong:falling}.} For a dissipative shock, $\Ws$ is negative and measures the energy lost per unit time by the internal frictions that hamper the shock as it  goes by. While in our setting the force $\Force$ will be provided through the contact of the slime filament with the external world, $\Ws$ is of a constitutive nature, which needs to be further specified (see Section~\ref{sec:shock_dissipation}). Equations \eqref{eq:jump_conditions_linear_momentum} and \eqref{eq:jump_conditions_energy} express only the mechanical balances at a singular point. The former is also known as the \emph{Rankine-Hugoniot} jump condition for one-dimensional continua \cite[p.\,29]{antman:nonlinear}.\footnote{The reader is further referred to \cite{reilly:treatment,reilly:energetics} for a general thermodynamic theory of strings, which also features an additional jump condition for the entropy imbalance. A formulation of shock waves for general three-dimensional continua can also be found in Secs.~32 and 33 of \cite{gurtin:mechanics}.}

Interesting versions of the jump conditions in \eqref{eq:kinematic_compatibility_velocity}, \eqref{eq:jump_conditions_linear_momentum}, and \eqref{eq:jump_conditions_energy} above occur when the material constituting the filament is amorphously quiescent on one side of the shock. In this context, for definiteness, we shall refer to such a shock as \emph{external}, while the shock described so far will be referred to as \emph{internal}.
An external shock is an attempt at formalizing the notion of \emph{continually} imparted impacts introduced in the work of Cayley~\cite{cayley:class}; as such, it is more than just an internal shock with vanishing velocity on one side. At an external shock, mass is not conserved, as the filament is there in contact with a slime reservoir, where a shapeless deposit of matter serves as a supply of mass abruptly injected into the moving filament. More generally, the moving system receives from the external reservoir supplies of mass, linear momentum, and energy, which enter the corresponding balance laws. This concept will inspire our mathematical treatment of the extrusion process that takes place at the bacterium pore whence the slime filament is extruded.

\nigh{Assuming that at the pore, where mass is instantaneously ejected with velocity $\vel$ along the direction $\ut_0$, a shock is propagating with speed $\sspeed$, the same kinematic condition that led us to \eqref{eq:kinematic_compatibility_velocity} now requires that}
\begin{subequations}\label{eq:jump_external_plus}
\begin{equation}\label{eq:jump_external_plus_velocity}
\vel+\sspeed\ut_0=\zero.
\end{equation}
\nigh{Similarly, by applying the balance laws of linear momentum and energy to an arbitrarily small arc of the ejected filament near the pore, we obtain}
\begin{equation}\label{eq:jump_external_plus_momentum}
\tautnot+\lambda\sspeed\vel+\Force_0=\zero,
\end{equation}
\begin{equation}\label{eq:jump_external_plus_energy}
\tautnot\cdot\vel+\frac12\lambda\sspeed v^2+W_0=0,
\end{equation}
\end{subequations}
\nigh{where we have denoted by $\tau_0$ the filament's tension at the pore, and $\Force_0$ and $W_0$ are the appropriate supplies.}\footnote{In particular, equation \eqref{eq:jump_external_plus_velocity} is nothing but the statement that slime is ejected along the tangent to the present shape of the filament. This is a necessary boundary condition for the existence of a steady solution of the dynamics of the filament that preserves its shape in the bacterium rest frame.} Combining \eqref{eq:jump_external_plus_velocity}, \eqref{eq:jump_external_plus_momentum}, and \eqref{eq:jump_external_plus_energy}, we see that
\begin{equation}\label{eq:jump_external_plus_consequences}
\begin{split}
\vel=v\ut_0,&\qquad\sspeed=-v,\\
\Force_0=-(\tau_0-\lambda v^2)\ut_0,&\qquad W_0=-\left(\tau_0-\frac12\lambda v^2\right)v.
\end{split}
\end{equation}

Equation \eqref{eq:jump_external_plus_consequences}, in particular, allows us to interpret $\Force_0$ as the \emph{continuous-impact} force envisaged by Cayley \cite{cayley:class} to describe mechanical systems in which particles of infinitesimal mass are continuously taken into ``connexion''. An external shock is propagating backwards relative to the filament at the same scalar velocity as the material in the filament, so that the shock results steady in space. The complementary expressions for $W_0$ give the energy lost (or gained) by the filament in being set in motion instantaneously.\footnote{It should perhaps be recalled that equations \eqref{eq:kinematic_compatibility_velocity} through \eqref{eq:jump_external_plus_consequences} are all valid in the bacterium rest frame.}

\subsection{Shock Dissipation}\label{sec:shock_dissipation}
When the shock is internal, that is, the singular point is both followed and preceded by mass in motion, the shock dissipation $\Ws$ should depend only on the impact mechanism responsible for the abrupt change in velocity. As in \cite{virga:dissipative}, to posit a constitutive law for $\Ws$, we seek inspiration in the laws of impact which were already introduced in 1668 by Wallis and Wren~\cite{wallis:summary}, as recounted, for example in Whittaker's treatise \cite[p.\,234]{whittaker:treatise}.

When in a system of mass-points all impacts happen to be characterized by the same restitution coefficient $0\leqq e\leqq1$, the kinetic energy after a single impact decreases by $(1-e)/(1+e)$ times the kinetic energy of the \emph{lost} motion, the motion that would have been composed with the motion before the impact to reproduce the motion after the impact \cite[p.\,235]{whittaker:treatise}. By applying this law to the elementary transfer of mass through the shock suffered by a filament, interpreted as an internal impact, we justify setting
\begin{equation}\label{eq:W_s_definition}
\Ws:=-\frac12f\lambda |\sspeed|\jump{\vel}^2,
\end{equation}
where $0\leqq f\leqq1$ will be treated as a phenomenological parameter.\footnote{Letting for a moment $f:=(1-e)/(1+e)$, we note that for a \emph{plastic} impact, $e=0$ and $f=1$, whereas, for a \emph{perfectly elastic} impact, $e=1$ and $f=0$. However, in the absence of a microscopic mechanism illuminating the origin of $\Ws$, these correspondences are purely formal and $f$ remains a constitutive parameter of the filament.} In the ideal limit where $f\to0^+$, the shock is not dissipative. On the other hand, for $f=1$, the shock is maximally dissipative. In practice, for a slime filament $f$ should depend on the material that constitutes it.\footnote{It might be suggested that $\Ws$ could equally well be regarded as an external energy sink, instead of the energy lost in an internal impact. This interpretation could indeed provide a better justification in the present context for the explicit law posited in \eqref{eq:W_s_definition}, \nigh{as the impact model recalled in the text, if appropriate for chain links, may be a bit too strained for a slime filament.}}

\subsection{Inverted Catenary}\label{sec:catenary}
\begin{figure}[!h]
\centering
  \includegraphics[width=.4\linewidth]{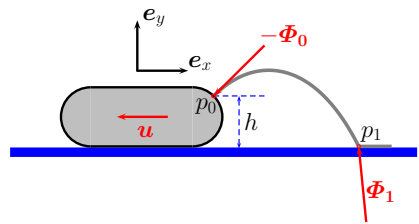}
  \caption{
  Snapshot of an idealized bacterium gliding over a fixed, flat substrate. In the Cartesian frame $(\ex,\ey)$, the gliding velocity is $\uv=-u\ex$. The points $p_0$ and $p_1$ represent the ejecting pore and the filament's foot, respectively; their difference in height in $h$. The force $\Force_0$ is applied in $p_0$ to the extruded filament, so that $-\Force_0$ is the force transmitted to the bacterium; its component along $\ex$ is the effective propelling force. $\Force_1$ is the linear momentum supply transferred to the internal shock in $p_1$ by the substrate.
  }
\label{fig:sketch}
\end{figure}

Before solving the balance equations for an ejected slime filament, we need to specify, albeit in an idealized fashion, the mechanisms at work at both pore and foot. Figure~\ref{fig:sketch} illustrates the view taken here. The points $p_0$ and $p_1$, which represent the pore and foot, respectively, are thought of as steady shocks, external the former and internal the latter, whose kinematic compatibility with the dynamic solution is still to be established.

The dynamics of a smooth arc of a slime filament is governed by equation \eqref{eq:smooth_momentum_balance}, while equations \eqref{eq:kinematic_compatibility_velocity} through \eqref{eq:jump_external_plus_consequences} are to be enforced at the two singular points identified above. We shall seek the solution to the problem within a special class, that of steady motions. To this end, we assume that the trajectory followed by the slime in the filament is invariable in time and that the spatial velocity field $\vel$ on it takes the form $\vel=v\ut$, with $v$ constant.

Projecting both sides of equation \eqref{eq:smooth_momentum_balance} along the tangent $\ut$, the principal normal $\n$ and the binormal $\bm{b}:=\ut\times\n$ to the filament's steady shape, we arrive at
\begin{equation}\label{eq:smooth_momentum_balance_components}
\tau'+f_t=0,\qquad(\lambda v^2-\tau)c=f_n,\qquad f_b=0,
\end{equation}
where $c$ is the shape's curvature and $f_t$, $f_n$, and $f_b$ are the components of $\force$ along $\ut$, $\n$, and $\bm{b}$, respectively (see also \cite{airy:mechanical}).

\begin{figure}[!h]
  \centering
  \subfigure[]{\includegraphics[width=.13\linewidth]{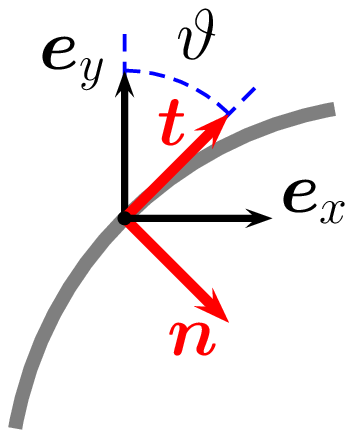}}\\
  \subfigure[]{\includegraphics[width=.15\linewidth]{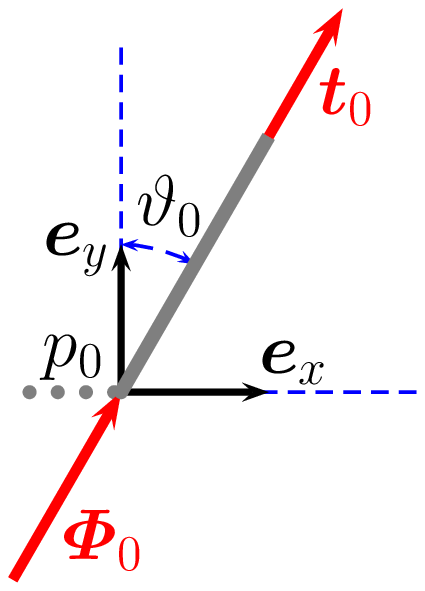}}
  \hspace{.1\linewidth}
  \subfigure[]{\includegraphics[width=.25\linewidth]{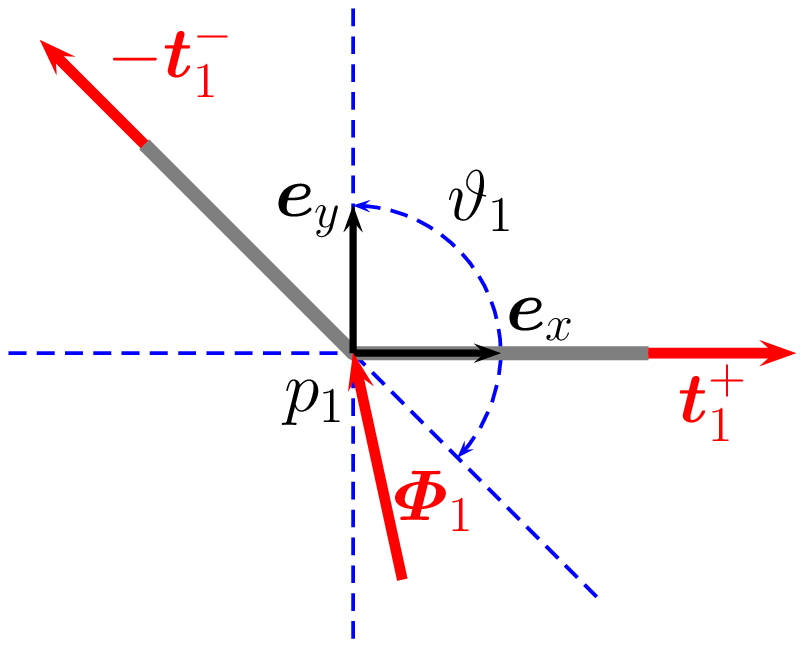}}
  \caption{
  Blowups of different significant portions of the steady shape of a slime filament:
  (a) Generic arc with the local, movable frame $(\ut,\n)$ and a fixed, Cartesian frame $(\ex,\ey)$. (b) Arc leaving the pore $p_0$; the dots represent the slime supply.  (c) Arc around the foot $p_1$. The different unit tangent vectors are described analytically by \eqref{eq:addition_3} and \eqref{eq:unit_tangents}. In our terminology, $p_0$ designates an external shock; a force $\Force_0$ is supplied there to the ejected slime filament, related to $\ut_0$ via \eqref{eq:jump_external_plus_consequences}. Similarly, $\Force_1$ is a realization of the momentum supply  $\Force$ featuring in \eqref{eq:jump_conditions_linear_momentum}.}
\label{fig:arc_and_jumps}
\end{figure}

Letting $\force$ lie in the $(x,y)$ plane, $f_b$ vanishes identically as long as the filament's shape lies in that plane as well. Figure~\ref{fig:arc_and_jumps}(a) describes a generic arc of the filament.
Denoting by $\vt$ the angle that $\ut$ makes with $\ey$, we can represent $\ut$ and $\n$ as
\begin{equation}\label{eq:movable_frame}
\ut=\sin\vt\,\ex+\cos\vt\,\ey,\quad\n=\cos\vt\,\ex-\sin\vt\,\ey,
\end{equation}
whence it follows that $c=\vt'$. Thus, as long as $c$ does not vanish, $\vt$ and $s$ can equally be employed to parameterize the filament's shape: in the setting described by Figs.~\ref{fig:sketch}, \ref{fig:arc_and_jumps}(b), and \ref{fig:arc_and_jumps}(c), $\tzero\leqq\vt\leqq\tone$. Expressing both $f_t$ and $f_n$ as functions of $\vt$, for $f_n\neq0$, we readily obtain from \eqref{eq:smooth_momentum_balance_components} that
\begin{equation}\label{eq:pre_shape_solution}
\ln|\lambda v^2-\tau|=\int\frac{f_t}{f_n}d\vt,\qquad c=\frac{f_n}{\lambda v^2-\tau}.
\end{equation}

If $\force=-\lambda g\ey$, where $g$ is the acceleration of gravity, then $f_t=-\lambda g\cos\vt$, $f_n=-\lambda g\sin\vt$, and \eqref{eq:pre_shape_solution} yields
\begin{equation}\label{eq:shape_solution}
\tau=\lambda v^2-\frac{a^2}{\sin\vt},\qquad c=\frac{\lambda g}{a^2}\sin^2\vt,
\end{equation}
where $a^2$ is a yet unknown, positive integration constant.\footnote{\nigh{Incidentally, neglecting both the filament's inertia and the gravity would result in setting equal to zero $\lambda$ and all components of $\force$ in \eqref{eq:smooth_momentum_balance_components}, thus leaving $\tau>0$ constant along a straight line and raising a geometric compatibility issue for the choice of the ejection angle $\tzero$, which would no longer be free.}} As already remarked in \cite[p.\,64]{walton:solutions}, the shape described by \eqref{eq:shape_solution} is an \emph{inverted} catenary. Moreover, for $\tau$ not to be negative somewhere, it suffices that $\tau_1:=\tau(\tone)\geqq0$, that is,
\begin{equation}\label{eq:a^2_inequality}
a^2\leqq\lambda v^2\sin\tone.
\end{equation}
By integrating in $\vt$, with the aid of \eqref{eq:shape_solution}, the equations
\begin{equation}\label{eq:shape_presolution_x_y}
\frac{dx}{d\vt}=\frac{\sin\vt}{c},\qquad\frac{dy}{d\vt}=\frac{\cos\vt}{c},
\end{equation}
which follow form \eqref{eq:movable_frame}, we arrive at
\begin{subequations}\label{eq:shape_solution_x_y}
\begin{equation}\label{eq:shape_solution_x}
x(\vt)=\frac{a^2}{\lambda g}\left(\ln\frac{1-\cos\vt}{\sin\vt}-\ln\frac{1-\cos\tzero}{\sin\tzero}\right),
\end{equation}
\begin{equation}\label{eq:shape_solution_y}
y(\vt)=\frac{a^2}{\lambda g}\left(\frac{1}{\sin\tzero}-\frac{1}{\sin\vt}\right),
\end{equation}
\end{subequations}
which parameterize the filament's steady shape in the Cartesian plane $(x,y)$ with origin at $p_0$. Likewise, the correspondence between $\vt$ and $s$ is expressed explicitly by
\begin{equation}\label{eq:shape_soltion_s_of_theta}
s(\vt)=\frac{a^2}{\lambda g}\left(\cot\tzero-\cot\vt\right).
\end{equation}

So far we have considered both the ejecting  velocity $v\geqq0$ and the ejecting angle $0\leqq\tzero\leqq\frac\pi2$ as parameters of the solution we seek. The solution of the balance equation for linear momentum along the ejected slime filament has identified two further parameters, $a^2$ and $\tone$, subject to the bound \eqref{eq:a^2_inequality}. In the next section, by use of appropriate boundary conditions, we shall resolve the shocks and devise a strategy to determine all four unknowns encountered here.

\section{Gliding Mechanics}\label{sec:gliding}
Here our theory is applied to describe the gliding of bacteria powered by slime extrusion. First, we introduce boundary conditions for which all balance equations of the theory can be solved. Then, we shall illustrate the solution thus found in a special case with a number of quantitative details.

\subsection{Motion Resolution}\label{sec:motion_resolution}
In the substrate rest frame, the gliding motion of the bacterium with constant velocity $\uv$ is governed by the equation
\begin{equation}\label{eq:addition_1}
-k_0\uv-(\Force_0\cdot\ex)\ex=\bm{0},
\end{equation}
which balances the viscous drag force, $-k_0\uv$, arising from the environment resistance and the effective propelling force, $\F=-(\Force_0\cdot\ex)\ex$, arising from the extruded slime filament. In \eqref{eq:addition_1}, $k_0$ is a viscosity coefficient characterizing the gliding bacterium and its environment, while $\F$ is determined by \eqref{eq:jump_external_plus_consequences} in the form
\begin{equation}\label{eq:addition_2}
\F=-(\lambda v^2-\tau_0)\sin\tzero\,\ex,
\end{equation}
where $v$ is the ejecting velocity and, in accord with \eqref{eq:movable_frame}, use has been made of the following representation for $\ut_0$,
\begin{equation}\label{eq:addition_3}
\ut_0=\sin\tzero\,\ex+\cos\tzero\,\ey.
\end{equation}

Denoting by
\begin{equation}\label{eq:addition_4}
F_0:=\lvtwo-\tau_0
\end{equation}
the propelling \emph{thrust}, which is  the total force felt by the bacterium, we rewrite \eqref{eq:addition_2} as $\F_0=-F_0\sin\tzero\,\ex$. In \eqref{eq:addition_4}, the tension $\tau_0$ of the filament at the pore is still to be determined; it will follow from the complete resolution of the shock equations. A comment is called upon by \eqref{eq:addition_4}: since $\tau_0\geqq0$, $F_0$ does not exceed $\lvtwo$, which according to \cite[p.\,29]{sommerfeld:mechanics} corresponds to the thrust that would be imparted by the jet propulsion of incoherent matter extruded at the rate $\lambda v$. This shows already that, however we shall determine $\tau_0$, the resulting thrust on the bacterium provided by an unbroken slime filament partially adhered to a rigid substrate will be less than that provided by pure jet propulsion. Letting $\uv=-u\ex$ and combining \eqref{eq:addition_1}, \eqref{eq:addition_2}, and \eqref{eq:shape_solution}, we arrive at
\begin{equation}\label{eq:addition_5}
u=\frac{a^2}{k_0},
\end{equation}
where $a^2$ is yet to be determined.

Now we resolve the two shocks shown in Fig.~\ref{fig:sketch} by applying equations \eqref{eq:kinematic_compatibility_velocity} through \eqref{eq:jump_conditions_energy} to $p_1$ and \eqref{eq:jump_external_plus_consequences} to $p_0$. In the bacterium rest frame the slime filament is seen to approach the substrate with velocity $\vel^-=v\ut^-_1$ before adhesion and to glide on the substrate with velocity $\vel^+=u\ut^+_1$ after adhesion; here the unit vectors $\ut^\pm_1$ are as shown in Fig.~\ref{fig:arc_and_jumps}(c), $v$ is the ejecting velocity, and $u$ is the gliding velocity.

By \eqref{eq:kinematic_compatibility_velocity}, we immediately conclude that $u=v$ and $\sspeed=-v$. Thus, kinematic compatibility requires that in a steady motion that preserves the filament's shape the bacterium gliding velocity $u$ (relative to the substrate) must coincide with the slime ejecting velocity $v$ (relative to the bacterium). Thus,  \eqref{eq:addition_5} is turned into an equation for the admissible ejecting velocity:
\begin{equation}\label{eq:addition_6}
v=\frac{a^2}{k_0},
\end{equation}
which depends on the filament's shape through the unknown constant $a^2$.

At $p_1$, equation \eqref{eq:jump_conditions_linear_momentum} can now be enforced in the form
\begin{equation}\label{eq:jump_conditions_linear_momentum_simplified}
\jump{(\tau-\lvtwo)\ut_1}+\Force_1=\zero.
\end{equation}
In a similar way, with the aid of \eqref{eq:W_s_definition}, at $p_1$ \eqref{eq:jump_conditions_energy} becomes
\begin{equation}\label{eq:jump_conditions_energy_simplified}
\jump{\tau}=\frac12 f\lvtwo\jump{\ut_1}^2.
\end{equation}
It is expedient recording here that, in accord with \eqref{eq:movable_frame}, the explicit expressions for the unit vectors $\ut_1^\pm$ featuring in \eqref{eq:jump_conditions_linear_momentum_simplified} and \eqref{eq:jump_conditions_energy_simplified} are
\begin{equation}\label{eq:unit_tangents}
\begin{split}
\ut_1^+&=\ex,\\
\ut_1^-&=\sin\tone\,\ex+\cos\tone\,\ey.
\end{split}
\end{equation}

While equation \eqref{eq:jump_conditions_linear_momentum_simplified}  determines the momentum supply $\Force_1$, the jump condition \eqref{eq:jump_conditions_energy_simplified} ties $a^2$ to $\tone$ via \eqref{eq:shape_solution}. Overall, there are four unknowns that need to be determined to identify completely the steady solution we seek here, namely, $\tone$, $v$, $a^2$, and $\tau^+$, where the latter designates the tension in the adhered filament, just after its foot $p_1$ (see Fig.~\ref{fig:sketch}). The bacterium equation of motion and the shock kinematic compatibility condition combined in \eqref{eq:addition_6} together with the jump condition \eqref{eq:jump_conditions_energy_simplified} provide only two equations: two others are missing.

One missing equation comes from the geometric condition
\begin{equation}\label{eq:geometric_condition}
y(\tone)=-h,
\end{equation}
which prescribes the total downfall of the filament (see Fig.~\ref{fig:sketch}). The other is the boundary condition that must be required at the filament's foot $p_1$ to reflect the contact mechanism with the substrate envisaged in the model. Here, to the purpose of showing that theoretically gliding is also  possible on smooth substrates, we shall assume that the reactive supply of linear momentum $\Force_1$ can only be directed vertically upwards, $\Force_1=\Phi_{1y}\ey$, with $\Phi_{y1}\geqq0$. Thus, the second missing equation will be
\begin{equation}\label{eq:addition_7}
\Phi_{1x}=\Force_1\cdot\ex=0.
\end{equation}

There are two compatibility conditions that a solution must meet to be acceptable: both concern the positivity of the tension $\tau$. One is \eqref{eq:a^2_inequality}, which amounts to require that $\tau_1\geqq0$, and the other is
\begin{equation}\label{eq:tension_positiveness_taum}
\tau^+\geqq0.
\end{equation}

To expedite the search for solutions and to retrace  more easily in them signs of universality, it is advisable to scale all lengths to $h$ and all velocities to
\begin{equation}\label{eq:addition_8}
v_0:=\frac{k_0}{\lambda},
\end{equation}
which represents a velocity characteristic of both the environment opposing the bacterium gliding and the material constituting the extruded slime. Thus, $v$ will be replaced by
\begin{equation}\label{eq:nu_definition}
\nu:=\frac{v}{v_0}.
\end{equation}

By \eqref{eq:nu_definition} and \eqref{eq:addition_8}, we readily rewrite \eqref{eq:addition_6} as
\begin{equation}\label{eq:addition_9}
a^2=\lambda\nu v^2_0,
\end{equation}
by which \eqref{eq:a^2_inequality} reduces to
\begin{equation}\label{eq:addition_10}
\nu\sin\tone\geqq1.
\end{equation}
By inserting \eqref{eq:addition_9} into \eqref{eq:geometric_condition}, we obtain the following expression for $\pi-\tzero\leqq\tone\leqq\pi$,
\begin{equation}\label{eq:addition_11}
\frac{1}{\sin\tone}=\frac{1}{\sin\tzero} +\frac{\eta^2}{2\nu},
\end{equation}
where
\begin{equation}\label{eq:addition_12}
\eta:=\frac{\sqrt{2gh}}{v_0},
\end{equation}
which represents the velocity acquired by any body falling from rest over the distance $h$ scaled to $v_0$, combines in a dimensionless parameter both $h$ and $v_0$. Similarly, using both \eqref{eq:jump_conditions_linear_momentum_simplified} and \eqref{eq:jump_conditions_energy_simplified}, we easily see that \eqref{eq:addition_7} amounts to
\begin{equation}\label{eq:addition_13}
\nu=\frac{1}{f\sin\tone},
\end{equation}
which makes \eqref{eq:addition_10} automatically satisfied since $0\leqq f\leqq1$.

Making use of \eqref{eq:addition_13} in \eqref{eq:addition_11}, we finally determine $\nu$ as
\begin{subequations}\label{eq:addition_14}
\begin{equation}\label{eq:addition_14a}
\nu=\nua:=\frac{1}{2f}\left(\frac{1}{\sin\tzero}+\sqrt{\frac{1}{\sin^2\tzero}+2f\eta^2}\right),
\end{equation}
in terms of which we express all other unknowns:
\begin{equation}\label{eq:addition_14b}
\tone=\pi-\arcsin\left(\frac{1}{f\nua}\right),
\end{equation}
\begin{equation}\label{eq:addition_14c}
\tau^+=\lambda v_0^2\nua(\nua-1),
\end{equation}
\begin{equation}\label{eq:addition_14d}
\Phi_{1y}=\lambda v_0^2\nua\sqrt{(f\nua)^2-1}.
\end{equation}
\end{subequations}
Equation \eqref{eq:addition_14c}, combined with \eqref{eq:addition_13}, shows in particular that $\tau^+$ satisfies the inequality \eqref{eq:tension_positiveness_taum}. Similarly, it follows from \eqref{eq:addition_14d} that $\Phi_{1y}\geqq0$. We record here for later use that by \eqref{eq:addition_9} the thrust in \eqref{eq:addition_4} can also be expressed as
\begin{equation}\label{eq:addition_15}
F_0=\frac{\lambda v^2_0\nua}{\sin\tzero}.
\end{equation}

\subsection{Energy Balance}\label{sec:energy_balance}
This is an active system, and so the energy dissipated in the gliding process must come from the bacterium itself. The theory developed here identifies with $W_0$ the rate at which such an active energy is produced to sustain the bacterium locomotion. In the bacterium rest frame, where the kinetic energy of the slime filament extruded between pore and foot is constant in time, balance of energy requires that
\begin{equation}\label{eq:energy_balance}
\dot{K}=W_0+\Ws+\Wg+\tau^+v,
\end{equation}
\nigh{where $K$ is the extra kinetic energy associated with the sliding motion of the filament on the substrate.}
In \eqref{eq:energy_balance}, $W_0$ is the power supply at the bacterium pore, given by \eqref{eq:jump_external_plus_consequences}, $\Ws$ is the concentrated power supply involved in the dissipative internal shock at the filament's foot, given by \eqref{eq:W_s_definition}, $\Wg$ is the power expended by the gravitational forces, and $\tau^+v$ is the power expended by the free end of the adhered filament (seen gliding with velocity $v$ in the bacterium rest frame). All these powers can be computed explicitly. To this end we found it expedient to scale them to $\lvthree$.

By combining \eqref{eq:jump_external_plus_consequences} with \eqref{eq:shape_solution} and replacing \eqref{eq:addition_9} with
\begin{equation}\label{eq:addition_16}
a^2=\lvtwo f\sin\tone,
\end{equation}
by use of both \eqref{eq:nu_definition} and \eqref{eq:addition_13}, we obtain
\begin{equation}\label{eq:addition_17}
W_0=\lvthree\left(f\frac{\sin\tone}{\sin\tzero}-\frac12\right).
\end{equation}
Directly from \eqref{eq:W_s_definition}, we have
\begin{equation}\label{eq:addition_17}
\Ws=-\lvthree f(1-\sin\tone).
\end{equation}
Moreover, it is easily seen (for example, in \cite{virga:dissipative}) that $\Wg=(\tau_0-\tau_1)v$, which by \eqref{eq:shape_solution} and \eqref{eq:addition_16} becomes
\begin{equation}\label{eq:addition_18}
\Wg=\lvthree f\left(1-\frac{\sin\tone}{\sin\tzero}\right).
\end{equation}
Likewise, by \eqref{eq:jump_conditions_energy_simplified} and \eqref{eq:addition_16}, we can write
\begin{equation}\label{eq:addition_19}
\tau^+v=\lvthree(1-f\sin\tone).
\end{equation}
Inserting equations \eqref{eq:addition_16} through \eqref{eq:addition_19} into \eqref{eq:addition_15}, we readily arrive at
\begin{equation}\label{eq:addition_20}
\dot{K}=\frac12\lvthree,
\end{equation}
which \nigh{is indeed justifiable by a direct computation. Considering the ever extending string of slime on the substrate, which moves at the speed $v$, we see that at time $t$ later it is carrying the extra kinetic energy $K=\frac12\lambda v^3t$ on account of being longer. By differentiating $K$ in time, we recover immediately \eqref{eq:addition_20}.}

In the following subsection we shall explore some quantitative aspects of this solution in the simpler case where $\tzero=\frac\pi2$ and $f$ and $\eta$ remain the only free parameters.

\subsection{Horizontal Ejection}\label{sec:horizontal}
\begin{figure}[!h]
\centering
  \includegraphics[width=.4\linewidth]{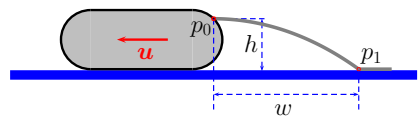}
  \caption{
  Sketch illustrating the solution in \eqref{eq:addition_14}. The filament's shape is drawn for the following choice of parameters: $\tzero=\frac\pi2$, $f=1$, $\eta=0.77$; the distance $w$ scaled to $h$ is $w/h=2.9$.
  }
\label{fig:sketch_horizontal}
\end{figure}

To illustrate the solution in \eqref{eq:addition_14}, here we set $\tzero=\frac\pi2$; the situation we envisage is sketched in Fig.~\ref{fig:sketch_horizontal}, where $w$ denotes the distance between the filament's foot $p_1$ and the projection onto the substrate of the ejecting pore $p_0$.

\begin{figure}[!h]
  \centering
  \subfigure[]{\includegraphics[width=.4\linewidth]{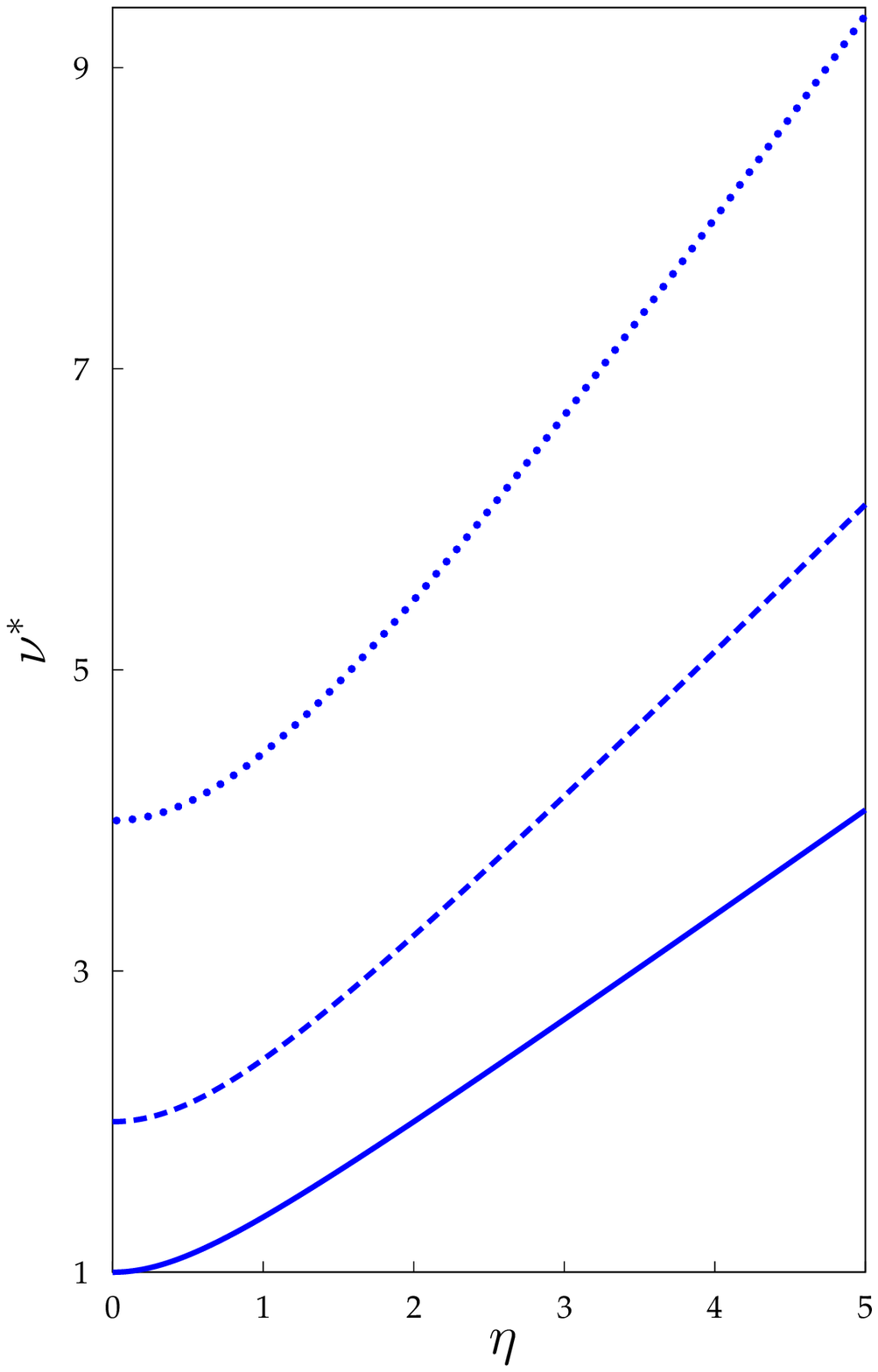}}
  \hspace{.05\linewidth}
  \subfigure[]{\includegraphics[width=.4\linewidth]{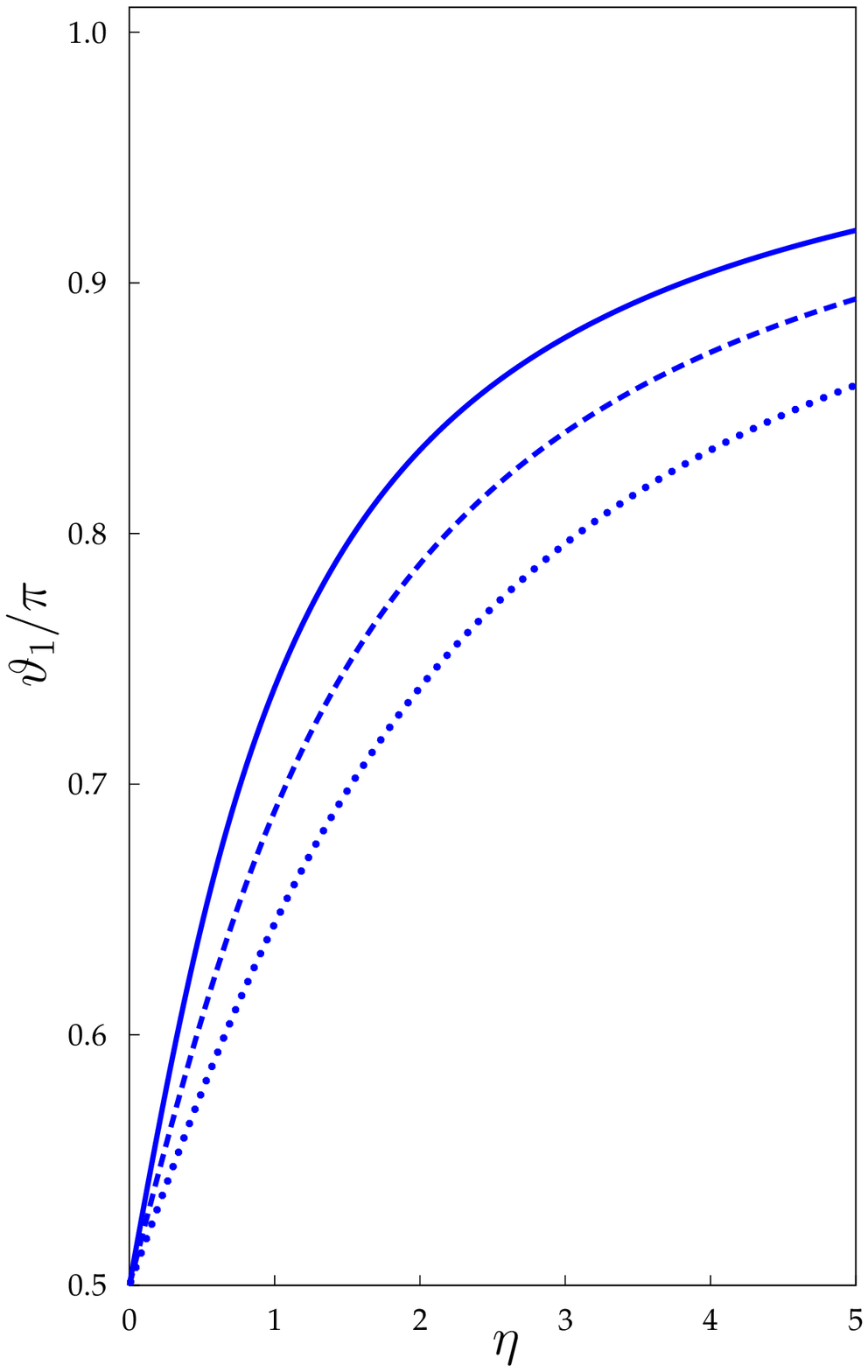}}\\
  \subfigure[]{\includegraphics[width=.4\linewidth]{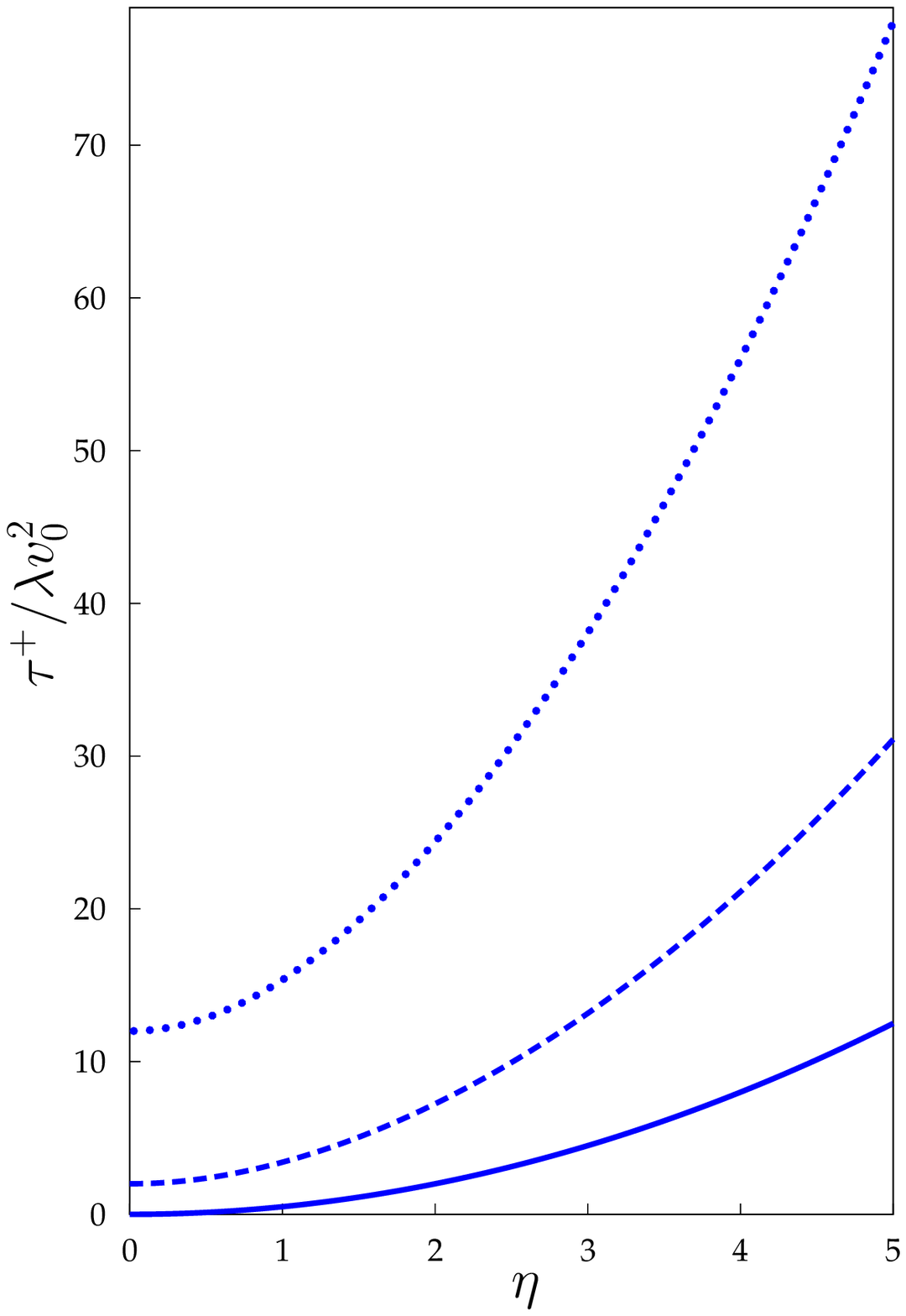}}
  \hspace{.05\linewidth}
  \subfigure[]{\includegraphics[width=.4\linewidth]{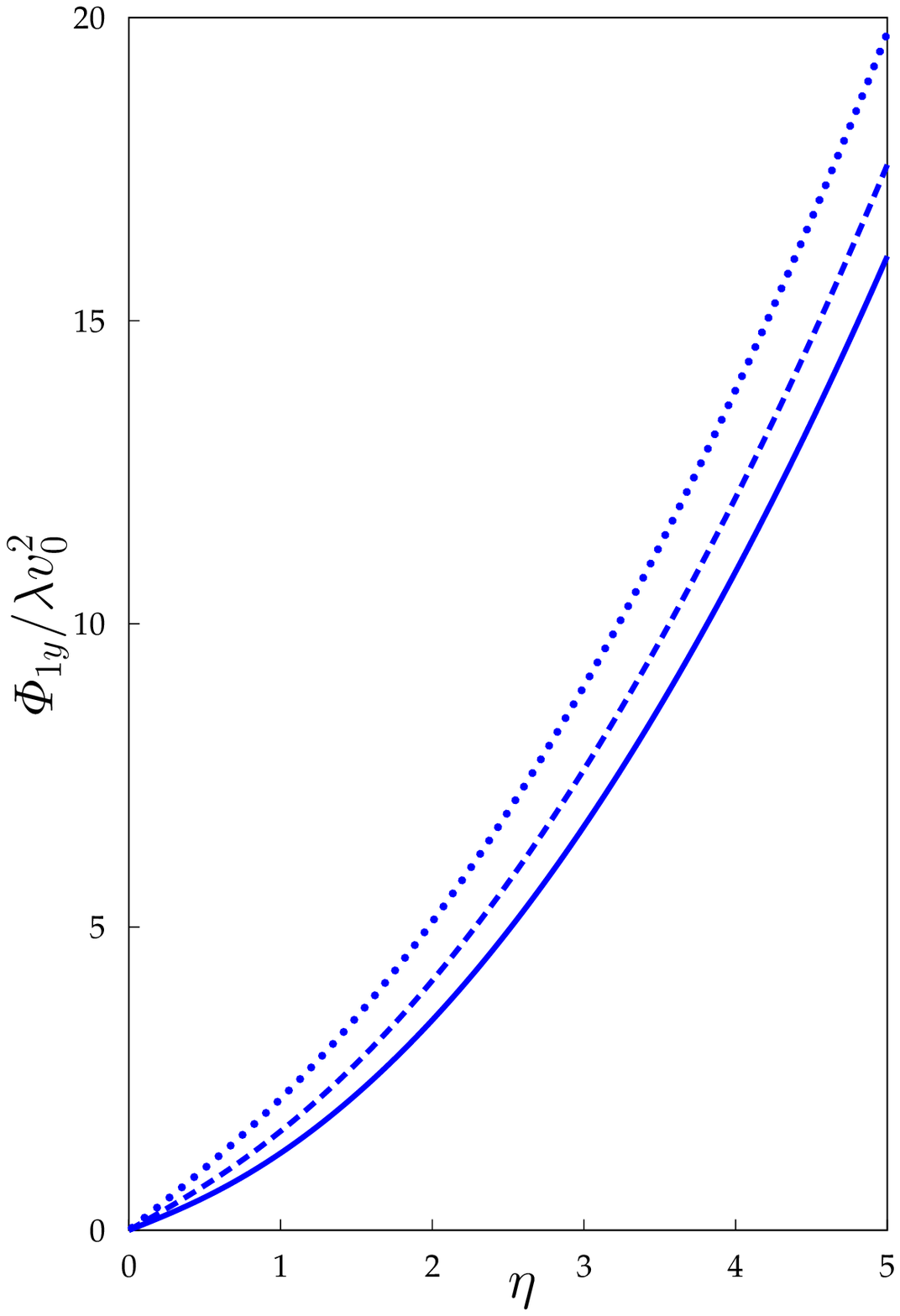}}
  \caption{
  The graphs of (a) $\nua$, (b) $\tone$, (c) $\tau^+$, and (d) $\Phi_{1y}$ as functions of $\eta$ for $f=\frac14$ (dotted lines), $f=\frac12$ (dashed lines), and $f=1$ (solid lines). \nigh{Here, as discussed in Section~\ref{sec:horizontal}, we have set $\tzero=\frac\pi2$.}}
\label{fig:tableau}
\end{figure}

The plots of $\nua$, $\tone$, $\tau^+$, and $\Phi_{1y}$ as functions of $\eta$ are shown in Fig.~\ref{fig:tableau} for three indicative values of $f$, namely, $\frac14$, $\frac12$, and $1$.
It should be noted that by \eqref{eq:addition_15}, for $\tzero=\frac\pi2$, $\nua$ also represents the thrust on the bacterium expressed in terms of $\lambda v_0^2$. All the graphs in Fig.~\ref{fig:tableau} describe increasing functions of $\eta$. Moreover, for a given $\eta$, increasing $f$ results in a decrease of $\nua$, $\tau^+$, and $\Phi_{1y}$, and in an increase of $\tone$. Thus, for a less dissipative shock, all the forces increase and so does also the gliding velocity, while the slime filament becomes steeper in approaching its foot.

\subsection{Neglecting Gravity}\label{sec:zero_gravity}
\nigh{The inverted catenary solution described in Section~\ref{sec:catenary} is characterized by gravity. One may sensibly argue that gravity is to play no role at the small length scales characteristic of bacteria. To put this objection into a quantitative perspective, we consider here what survives of our analysis in the limit as the dimensionless parameter $\eta$ introduced in \eqref{eq:addition_12} tends to zero; we interpret this to be either the zero gravity limit or the zero thickness limit, in view of the way both $g$ and $h$ feature in \eqref{eq:addition_12}.
By use of \eqref{eq:addition_14}, it is a simple matter to show that for $\eta\to0$ $\nua$, $\tone$, $\tau^+$, and $\Phi_{1y}$ are respectively delivered by
\begin{subequations}\label{eq:zero_gravity_limit}
\begin{equation}\label{eq:zero_gravity_limit_a}
\nnot{\nua}=\frac{1}{f\sin\tzero},
\end{equation}
\begin{equation}\label{eq:zero_gravity_limit_b}
\nnot{\tone}=\tzero,
\end{equation}
\begin{equation}\label{eq:zero_gravity_limit_c}
\nnot{\tau^+}=\frac{\lambda v_0^2}{f^2\sin\tzero}(1-f\sin\tzero),
\end{equation}
\begin{equation}\label{eq:zero_gravity_limit_d}
\nnot{\Phi_{1y}}=\frac{\lambda v_0^2}{f\sin^2\tzero}\cos\tzero.
\end{equation}
\end{subequations}
Correspondingly, by \eqref{eq:addition_12}, \eqref{eq:addition_9}, and \eqref{eq:shape_solution_y}, for any $\eta$ the function $y(\vt)$ that describes the inverted catenary can be recast in the form
\begin{equation}\label{eq:y_catenary_recast}
\frac{y(\vt)}{h}=\frac{1-\frac{\sin\tzero}{\sin\vt}}{\frac{\sin\tzero}{\sin\tone}-1}.
\end{equation}
Since by \eqref{eq:zero_gravity_limit_b} $\tone$ tends to $\tzero$ as $\eta$ tends to $0$, \eqref{eq:y_catenary_recast} shows that in that limit the inverted catenary solution makes no sense, and so the filament's shape remains undetermined (as was perhaps to be expected). Nevertheless, all formulas \eqref{eq:zero_gravity_limit} remain perfectly valid and carry a definite mechanical meaning, which is presumably best appropriate at bacterial length scales.
}

\section{Conclusion}\label{sec:conclusion}
We have proposed a mathematical model to describe the mechanics of gliding, a means utilized by some bacteria to move on a rigid substrate. The theory attempted to prove the propelling ability of the slime filaments extruded by some cells, such as myxobacteria and cyanobacteria, which are known to adopt gliding as a locomotion mechanism.

Seen in the bacterium rest frame, an extruded slime filament was treated as a flexible, inextensible string with uniform linear mass density, flowing along its own shape and meeting the substrate at a kink, which we called the \emph{foot}, where a dissipative internal shock travels backwards in the string reference configuration to remain still relative to the moving bacterium. The extruding pore was instead treated as an external shock, as there new mass is continuously supplied to the moving filament together with linear momentum and energy. This distinction between internal and external shocks is taken from \cite{virga:dissipative}; the mathematical formalism put forward there was also recalled here to make this paper self-contained and adapted to an ideal bacterium gliding over a flat, rigid substrate.

We proved that the kinetic compatibility condition for shock propagation requires the gliding velocity (in the substrate rest frame) to equal the extrusion velocity (in the bacterium rest frame), a result that seems to be supported by some experimental evidence \cite{wolgemuth:biomechanics}. \nigh{Essentially, this is a consequence of our assumption on the filament's inextensibility, in favour of which we were not able to produce any direct experimental validation. Perhaps, we may consider the observed coincidence between gliding and extruding velocities as an indirect one.} We also determined completely the bacterium gliding motion by solving explicitly a non-linear mathematical problem. We obtained both the velocity compatible with the assumption of uniform gliding and the propelling thrust on the bacterium in terms of a few parameters.  Though not numerous, these latter are not directly accessible, so that we failed to propose any reliable estimate of the mechanical quantities involved in the bacterium motion. \nigh{In our analysis, gravity was not neglected, whereas both viscous forces on the extruded filament and its bending rigidity were. As for the first assumption, which might be considered inappropriate at the typical length scales of bacteria, we proved that in the zero gravity limit only the specific inverted catenary structure of the solution does not survive (as was perhaps to be expected), whereas all the mechanical quantities we were interested in have a definite limit, so that gravity appears here as an analytic regularizing device which just makes the filament's shape definite. As for the other two assumptions, I simply plead guilty of having no better justification than the simplicity they afford to the analysis.}\footnote{\nigh{It was already shown in \cite{airy:mechanical} that viscous forces make the simple catenary problem far more complicated, though still solvable analytically in some special cases.}}

At this stage, ours remains just a viable mathematical hypothesis on the role of slime extrusion in bacteria gliding. We were contented with proving theoretically that continuously extruded slime filaments exert a thrust on the bacterium through their footing on the substrate, for which we can afford a precise quantitative description. Whether this would contribute to clarify the mystery of bacteria gliding still remains to be seen.

\section*{Acknowledgment}
\nigh{I wish to thank both anonymous Referees, whose remarks and suggestions helped me putting my work in a better perspective. I am, in particular, indebted to one of them for having prompted me to write Section~\ref{sec:zero_gravity} about the zero gravity limit and for correcting a shortcoming in a previous version of Section~\ref{sec:energy_balance}; the lines following \eqref{eq:addition_20} are essentially taken from his (or her) report.}


\begin{thebibliography}{10}

\bibitem{wolgemuth:biomechanics}
Wolgemuth CW. 2012 Biomechanics of cell motility. \emph{In} EH~Egelman (Ed.),
  \emph{Comprehensive Biophysics}, Amsterdam: Elsevier, 168--193.

\bibitem{nan:uncovering}
Nan B, Zusman DR. 2011 Uncovering the mystery of gliding motility in the
  myxobacteria. \emph{Annu. Rev. Genet.} \textbf{45}, 21--39.

\bibitem{jarosch:gliding}
Jarosch R. 1962 Gliding. \emph{In} RA~Lewin (Ed.), \emph{Biochemistry and
  physiology of algae}, New York, 573--581.

\bibitem{walsby:mucilage}
Walsby A. 1968 Mucilage secretion and the movements of blue-green algae.
  \emph{Protoplasma} \textbf{65}, 223--238.

\bibitem{reichenbach:taxonomy}
Reichenbach H. 1981 The taxonomy of the gliding bacteria. \emph{Annu. Rev.
  Microbiol.} \textbf{35}, 339--364.

\bibitem{fritsch:structure}
Fritsch FE. 1945 \emph{The structure and reproduction of the algae}, volume~2.
  Cambridge: Cambridge University Press, 4th edition.

\bibitem{spagnolie:jet}
Spagnolie SE, Lauga E. 2010 Jet propulsion without inertia. \emph{Phys. Fluids}
  \textbf{22}, 081902.

\bibitem{burchard:gliding_prokaryotes}
Burchard RP. 1981 Gliding motility of prokaryotes: ultrastructure, physiology,
  and genetics. \emph{Annu. Rev. Microbiol.} \textbf{35}, 497--529.

\bibitem{hoiczyk:gliding}
Hoiczyk E. 2000 Gliding motility in cyanobacteria: observations and possible
  explanations. \emph{Arch. Microbiol.} \textbf{174}, 11--17.

\bibitem{spormann:gliding}
A~Spormann DK. 1995 Gliding movements in \emph{Myxococcus xanthus}. \emph{J.
  Bacteriol.} \textbf{177}, 5846--5852.

\bibitem{mauriello:gliding}
Mauriello EM, Mignot T, Yang Z, Zusman DR. 2010 Gliding motility revisited: How
  do the myxobacteria move without flagella? \emph{Microbiol. Mol. Biol. Rev.}
  \textbf{74}, 229--249.

\bibitem{koch:social}
Koch A, White D. 1998 The social lifestyle of myxobacteria. \emph{Bioessays}
  \textbf{20}, 1030--1038.

\bibitem{shimkets:social}
Shimkets LJ. 1990 Social and developmental biology of the myxobacteria.
  \emph{Microbiol. Rev.} \textbf{54}, 473--501.

\bibitem{kaiser:coupling}
Kaiser D. 2003 Coupling cell movement to multicellular development in
  myxobacteria. \emph{Nat. Rev. Microbiol.} \textbf{1}, 45--54.

\bibitem{igoshin:waves}
Igoshin OA, Welch R, Kaiser D, Oster G. 2004 Waves and aggregation patterns in
  myxobacteria. \emph{Proc. Natl. Acad. Sci. USA} \textbf{101}, 4256--4261.

\bibitem{bradley:function}
Bradley DE. 1980 A function of \emph{Pseudomonas aeruginosa} {PAO} polar pili:
  twitching motility. \emph{Can. J. Microbiol.} \textbf{26}, 146--154.

\bibitem{hoiczyk:genetics}
Hoiczyk E, Baumeister W. 1998 The junctional pore complex, a prokaryotic
  secretion organelle, is the molecular motor underlying gliding motility in
  cyanobacteria. \emph{Curr. Biol.} \textbf{8}, 1161--1168.

\bibitem{jahn:polyangiden}
Jahn E. 1924 Die polyangiden. \emph{In} G~Borntraeger (Ed.), \emph{Beitrage zur
  botanischen Protistologie}, Leipzig.

\bibitem{kuhlwein:weitere}
Kuhlwein H. 1953 Weitere untersuchungen an myxobacterien. \emph{Arch.
  Mikrobiol.} \textbf{19}, 365--371.

\bibitem{stanier:elasticotaxis}
Stanier R. 1942 Elasticotaxis in myxobacteria. \emph{J. Bacteriol.}
  \textbf{44}, 405--412.

\bibitem{fontes:myxococcus}
Fontes KD M. 1999 \emph{Myxococcus} cells respond to elastic forces in their
  substrate. \emph{Proc. Natl. Acad. Sci. USA} \textbf{96}, 8052--8057.

\bibitem{wolgemuth:how}
Wolgemuth C, Hoiczyk E, Kaiser D, Oster G. 2002 How myxobacteria glide.
  \emph{Curr. Biol.} \textbf{12}, 369--377.

\bibitem{wolgemuth:junctional}
Wolgemuth CW, Oster G. 2004 The junctional pore complex and the propulsion of
  bacterial cells. \emph{J. Mol. Microbiol. Biotechnol.} \textbf{7}, 72--77.

\bibitem{sun:effect}
Sun H, Yang Z, Shi W. 1999 Effect of cellular filamentation on adventurous and
  social gliding motility of \emph{Myxococcus xanthus}. \emph{Proc. Natl. Acad.
  Sci. USA} \textbf{96}, 15178--15183.

\bibitem{sliusarenko:motors}
Sliusarenko O, Zusman DR, Oster G. 2007 The motors powering {A}-motility in
  \emph{Myxococcus xanthus} are distributed along the cell body. \emph{J.
  Bacteriol.} \textbf{189}, 7920--7921.

\bibitem{mignot:evidence}
Mignot T, Shaevitz JW, Hartzell PL, Zusman DR. 2007 Evidence that focal
  adhesion complexes power bacterial gliding motility. \emph{Science}
  \textbf{315}, 853--856.

\bibitem{luciano:emergence}
Luciano J, Agrebi R, {Le Gall} AV, Wartel M, Fiegna F, Ducret A,
  Brochier-Armanet C, Mignot T. 2011 Emergence and modular evolution of a novel
  motility machinery in bacteria. \emph{PLoS Genet.} \textbf{9}, e1002268.

\bibitem{ducret:wet}
Ducret A, Valignat MP, Mouhamar F, Mignot T, Theodoly O. 2012
  Wet-surface-enhanced ellipsometric contrast microscopy identifies slime as a
  major adhesion factor during bacterial surface motility. \emph{Proc. Natl.
  Acad. Sci. USA} \textbf{109}, 10036--10041.

\bibitem{jarrell:surprisingly}
Jarrell KF, McBride MJ. 2008 The surprisingly diverse ways that prokaryotes
  move. \emph{Nat. Rev. Micro.} \textbf{6}, 466--476.

\bibitem{koch:characterization}
Koch MK, Hoiczyk E. 2013 Characterization of myxobacterial {A}-motility:
  insights from microcinematographic observations. \emph{J. Basic Microbiol.}
  \textbf{53}, 785--791.

\bibitem{nan:flagella}
Nan B, Bandaria JN, Moghtaderi A, Sun IH, Yildiz A, Zusman DR. 2013 Flagella
  stator homologs function as motors for myxobacterial gliding motility by
  moving in helical trajectories. \emph{Proc. Natl. Acad. Sci. USA}
  \textbf{110}, E1508--E1513.

\bibitem{hodgkin:junctional}
Hodgkin J, Kaiser D. 1979 Genetics of gliding motility in \emph{Myxococcus
  xanthus} (myxobacterales): two gene systems control movement. \emph{Mol. Gen.
  Genet.} \textbf{171}, 177--191.

\bibitem{guglielmi:structure}
Guglielmi G, Cohen-Bazire G. 1982 Structure et distribution des pores et des
  perforations de l'enveloppe de peptidoglycane chez quelques cyanobact\'eries.
  \emph{Protistologica} \textbf{18}, 151.

\bibitem{hoiczyk:envelope}
Hoiczyk E, Baumeister W. 1995 Envelope structure of four gliding filamentous
  cyanobacteria. \emph{J. Bacteriol.} \textbf{177}, 2387--2395.

\bibitem{dhahri:in-situ}
Dhahri S, Ramonda M, Marli\`ere C. 2013 In-situ determination of the mechanical
  properties of gliding or non-motile bacteria by atomic force microscopy under
  physiological conditions without immobilization. \emph{PLoS ONE} \textbf{8},
  e61663.

\bibitem{virga:dissipative}
Virga EG. 2014 Dissipative shocks in a chain fountain. \emph{Phys. Rev. E}
  \textbf{89}, 053201.

\bibitem{mould:self}
Mould S. 2013 Self siphoning beads.
  \url{http://stevemould.com/siphoning-beads/}.

\bibitem{hanna:instability}
Hanna JA, King H. 2011 An instability in a straightening chain.
  \emph{arXiv:1110.2360 [physics.flu-dyn]}
  \url{http://arxiv.org/abs/1110.2360}.

\bibitem{hanna:slack}
Hanna JA, Santangelo CD. 2012 Slack dynamics on an unfurling string.
  \emph{Phys. Rev. Lett.} \textbf{109}, 134301.

\bibitem{biggins:understanding}
Biggins JS, Warner M. 2014 Understanding the chain fountain. \emph{Proc. Roy.
  Soc. London A} \textbf{470}, 20130689.

\bibitem{biggins:growth}
Biggins JS. 2014 Growth and shape of a chain fountain. \emph{EPL} \textbf{106},
  44001.

\bibitem{reilly:treatment}
O'Reilly OM, Varadi PC. 1999 A treatment of shocks in one-dimensional
  thermomechanical media. \emph{Continuum Mech. Thermodyn.} \textbf{11},
  339--352.

\bibitem{green:thermodynamics}
Green AE, Naghdi PM. 1977 On thermodynamics and the nature of the second law.
  \emph{Proc. Roy. Soc. London A} \textbf{357}, 253--270.

\bibitem{green:note}
Green AE, Naghdi PM. 1977 A note on thermodynamics of constrained materials.
  \emph{J. Appl. Mech.} \textbf{44}, 787--788.

\bibitem{green:derivation}
Green A, Naghdi P. 1978 A derivation of jump condition for entropy in
  thermomechanics. \emph{J. Elast.} \textbf{8}, 179--182.

\bibitem{green:thermal}
Green A, Naghdi P. 1979 On thermal effects in the theory of rods. \emph{Int. J.
  Solids Structures} \textbf{15}, 829--853.

\bibitem{villaggio:mathematical}
Villaggio P. 1997 \emph{Mathematical Models for elastic structures}. Cambridge:
  Cambridge University Press.

\bibitem{reilly:energetics}
O'Reilly OM, Varadi PC. 2003 On energetics and conservations for strings in the
  presence of singular sources of momentum and energy. \emph{Acta Mech.}
  \textbf{165}, 27--45.

\bibitem{sommerfeld:mechanics}
Sommerfeld A. 1964 \emph{Mechanics}, volume~1 of \emph{Lectures on Theoretical
  Physics}. New York: Academic Press.

\bibitem{muller:history}
M\"uller I. 2007 \emph{A History of Thermodynamics}. Berlin: Springer-Verlag.

\bibitem{steiner:equations}
Steiner W, Troger H. 1995 On the equations of motion of the folded inextensible
  string. \emph{Z. angew. Math. Phys. (ZAMP)} \textbf{46}, 960--970.

\bibitem{wong:falling}
Wong CW, Yasui K. 2006 Falling chains. \emph{Am. J. Phys.} \textbf{74},
  490--496.

\bibitem{antman:nonlinear}
Antman SS. 1995 \emph{Nonlinear Problems of Elasticity}, volume 107 of
  \emph{Applied Mathematical Sciences}. New York: Springer.

\bibitem{gurtin:mechanics}
Gurtin ME, Fried E, Anand L. 2010 \emph{The Mechanics and Thermodynamics of
  Contiuna}. Cambridge: Cambridge University Press.

\bibitem{cayley:class}
Cayley A. 1856--1857 On a class of dynamical problems. \emph{Proc. Roy. Soc.
  London} \textbf{8}, 506--511.

\bibitem{wallis:summary}
Wallis J, Wren C. 1668 A summary account of the general laws of motion by dr.
  john wallis, and dr. christopher wren. \emph{Phil. Trans. Roy. Soc. London}
  \textbf{3}, 864--868.

\bibitem{whittaker:treatise}
Whittaker ET. 1937 \emph{A Treatise on the Analytical Dynamics of Particles and
  Rigid Bodies}. Cambridge: Cambridge University Press, 4th edition. Reissued
  with Forward in the Cambridge Mathematical Library Series 1988.

\bibitem{airy:mechanical}
Airy GB. 1858 On the mechanical conditions of the deposit of a submarine cable.
  \emph{Phil. Mag. S. 4} \textbf{16}, 1--18.

\bibitem{walton:solutions}
Walton W, Mackenzie C. 2005 \emph{Solutions of the Problems and Riders Proposed
  in the Senate-House Examination for 1854 by the Moderators and Examiners}.
  Boston: Adamant Media Corporation. Replica of 1854 edition by Macmillan and
  Co., Cambridge.

\end{thebibliography}


\end{document}